
\input phyzzx
\overfullrule=0pt

\def\gl{\lambda}
\def\p{\partial}
\def\phip{\p_+\phi}
\def\phim{\p_-\phi}

\def\rhop{\p_+\rho}
\def\rhom{\p_-\rho}
\def\rhopm{\p_+\p_-\rho}
\def\chip{\p_+\chi}
\def\chim{\p_-\chi}
\def\chipm{\p_+\p_-\chi}
\def\omp{\p_+\Omega}
\def\omm{\p_-\Omega}

\rightline {SU-ITP-92-17}
\rightline {June 1992}
\bigskip\bigskip
\title{The Endpoint of Hawking Radiation\foot{Work
supported in
part by NSF grant PHY89-17438.}}

\vfill
\author{Jorge G. Russo,\foot{INFN associate,
Sezione di Trieste, supported by an INFN fellowship.}
Leonard Susskind and L\'arus Thorlacius}
\bigskip
\address{ Department of Physics \break Stanford University,
            Stanford, CA 94305}
\vfill

\abstract
\singlespace
The formation and semi-classical evaporation of two-dimensional black
holes is studied in an exactly solvable model.  Above a certain
threshold energy flux, collapsing matter forms a singularity inside
an apparent horizon.  As the black hole evaporates the apparent
horizon recedes and meets the singularity in a finite proper time.
The singularity emerges naked and future evolution of the geometry
requires boundary conditions to be imposed there.  There is a natural
choice of boundary conditions which match the evaporated black hole
solution onto the linear dilaton vacuum.  Below the threshold energy
flux no horizon forms and boundary conditions can be imposed where
infalling matter is
reflected from a time-like naked singularity.  All information is
recovered at spatial infinity in this case.

\vfill\endpage

\REF\cghs{C.~G.~Callan, S.~B.~Giddings, J.~A.~Harvey and
A.~Strominger
\journal Phys. Rev. & D45 (92) R1005.}

\REF\bddo{T.~Banks, A.~Dabholkar, M.~R.~Douglas and M.~O'Loughlin,
\journal Phys. Rev. & D45 (92) 3607.}

\REF\rst{J.~G.~Russo, L.~Susskind and L.~Thorlacius, {\it Black
Hole Evaporation in 1+1 Dimensions}, Stanford University preprint,
SU-ITP-92-4, January 1992.}

\REF\bghs{B.~Birnir, S.~B.~Giddings, J.~A.~Harvey and A.~Strominger,
{\it Quantum Black Holes}, preprint UCSB-TH-92-08, EFI-92-16,
March 1992.}

\REF\haw{S.~W.~Hawking, {\it Evaporation of Two Dimensional Black
Holes}, Caltech preprint, CALT-68-\#1, March 1992.}

\REF\lslt{L.~Susskind and L.~Thorlacius, {\it Hawking Radiation and
Back-Reaction}, SU-ITP-92-12, hepth@xxx/9203054, March 1992.}

\REF\strom{A.~Strominger, {\it Fadeev-Popov Ghosts and 1+1
Dimensional Black
Hole Evaporation}, UC Santa Barbara preprint, UCSB-TH-92-18,
hepth@xxx/9205028,
May 1992.}

\REF\jrat{J.~G.~Russo and A.~A.~Tseytlin, {\it Scalar Tensor Quantum
Gravity in Two-Dimensions}, Stanford University preprint,
SU-ITP-92-2, hepth@xxx/9201021, January 1992.}

\REF\bilcal{A.~Bilal and C.~G.~Callan, {\it Liouville Models of Black
Hole
Evaporation}, Princeton University preprint, PUPT-1320,
hepth@xxx/9205089, May
1992.}

\REF\dealw{S.~P.~de~Alwis, {\it Quantization of a Theory of 2d
Dilaton
Gravity}, University of Colorado preprint, COLO-HEP-280,
hepth@xxx/9205069, May
1992; {\it Black Hole Physics from Liouville Theory}, COLO-HEP-284,
hepth@xxx/9206020, June 1992.}

\REF\hawtalk{S.~W.~Hawking, talk delivered at SISSA, Trieste, April
1992.}

\REF\andy{A.~Strominger, private communication, June 1992.}

\REF\cgclt{C.~G.~Callan and L.~Thorlacius
\journal Nucl. Phys. & B319 (89) 133
\journal Nucl. Phys. & B329 (90) 117.}

\REF\pst{A.~Peet, L.~Susskind and L.~Thorlacius, {\it Information
Loss and Anomalous Scattering}, Stanford University preprint,
SU-ITP-92-16, June 1992.}

\chapter{Introduction}

Recent months have seen a lot of activity in the study of quantum
effects on black holes.  For this purpose, Callan, Giddings, Harvey
and Strominger [\cghs] proposed a simple two-dimensional model
involving gravity coupled to a dilaton and conformal matter fields.
The model has classical solutions which describe the formation of
black holes and enables a semi-classical treatment of Hawking
radiation and its back-reaction on the geometry.  This has been
developed further by a number of authors [\bddo -\dealw].  The
semi-classical equations of the CGHS-model have not been solved in
closed form but recently Bilal and Callan [\bilcal] and de Alwis
[\dealw] have shown how the original model can be modified to allow
explicit construction of exact quantum black hole solutions.
Astutely chosen field redefinitions allow the modified theory
to be written as a Liouville model and the semi-classical equations
are straightforwardly solved in that form.

In this paper we will study a variation on this theme.  Rather than
modifying the dilaton potential to achieve a solvable field theory,
as was done in [\jrat -\dealw], we instead change the kinetic
term.  This change simplifies somewhat the field redefinitions which
take the model into a Liouville theory and also allows us to identify
the vacuum configuration in a straightforward fashion.  While the
emphasis in [\jrat -\dealw] was on achieving consistent conformal
field theories our goal here is limited to modifying the
CGHS-equations to make them exactly solvable, and then to study the
physics of the resulting solutions.  The complete quantization of
this system is outside the scope of this paper.

The classical action of the original CGHS-model is
$$
S_0 = {1\over 2\pi} \int d^2x\>\sqrt {-g} \bigl[
  e^{-2\phi}(R+4(\nabla \phi )^2
+4\gl ^2)-{1\over 2}\sum_{i=1}^N (\nabla f_i)^2 \bigr]\> .
\eqn\lagr
$$
In conformal gauge, $g_{++}=g_{--}=0$, $g_{+-}=-{1\over 2}
e^{2\rho}$, this becomes
$$
S_0 = {1\over \pi} \int d^2x\>
\bigl[e^{-2\phi}(2\rhopm - 4 \phip \phim
+\gl ^2 e^{2\rho})
-{1\over 2}\sum_{i=1}^N \p_+ f_i \p_- f_i \bigr]\> .
\eqn\first
$$
In addition to the equations of motion of $\rho$, $\phi$ and $f_i$ we
have to impose two constraints corresponding to the equations of
motion of the vanishing metric components,
$$
0 = e^{-2\phi}(4 \p_\pm\rho \p_\pm\phi - 2\p_\pm^2\phi)
      + {1\over 2} \sum_{i=1}^N \p_\pm f_i \p_\pm f_i \> .
\eqn\second
$$
The classical action has a useful symmetry,
$$\delta\phi=\delta\rho=\epsilon e^{2\phi}\> ,
\eqn\third
$$
where $\epsilon$ is infinitesimal.  The associated conserved current
is
$$
j^\mu = \partial ^\mu (\phi - \rho)\> ,
\eqn\fourth
$$
and the conservation equation is
$$
\partial _\mu \partial ^\mu (\phi - \rho) = 0 \> .
\eqn\fifth
$$
This fact allows one to choose a special conformal gauge in which
$\rho=\phi$.  It turns out to be convenient to preserve this simple
form of the current at the one-loop level.  This enables us to use
the special conformal gauge when solving the semi-classical
equations.  We can always add covariant terms to the definition of
our model for this purpose.

Let us now consider one-loop quantum corrections.  The matter fields
contribute the familiar conformal anomaly term and we also add a
local and covariant term to preserve the simple form of the current
\fourth .\foot{The symmetry transformation \third\ will receive
quantum
corrections
$\delta\phi=\delta\rho=\epsilon e^{2\phi}/ (1-{\kappa\over
4}e^{2\phi})$
but the conserved current \fourth\ remains unchanged.}  One obtains
the following effective action,
$$\eqalign{
S =& S_0 - {\kappa\over 8\pi} \int d^2x\>
\sqrt{-g} \bigl[R\,{1\over \nabla^2}\,R
- 2\phi \,R \bigr] \> ,  \cr
=& S_0 - {\kappa\over \pi} \int d^2x\>
[\rhop\rhom +\phi\rhopm]
\> , \cr}
\eqn\sixth
$$
where $\kappa ={N\over 12}$, and the constraints become
$$\eqalign{
0 = (e^{-2\phi}-{\kappa\over 4})\,
(4 \p_\pm\rho &\p_\pm\phi - 2\p_\pm^2\phi)
 + {1\over 2} \sum_{i=1}^N \p_\pm f_i \p_\pm f_i \cr
 &- \kappa (\p_\pm\rho \p_\pm\rho - \p_\pm^2\rho
     +t_\pm) \> . \cr}
\eqn\seventh
$$
The functions $t_\pm(x^\pm)$ reflect the non-local nature of the
anomaly and are determined by boundary conditions [\cghs].  The
contribution to the constraints from our extra term in \sixth\
vanishes on classical solutions so it will not affect the rate of
Hawking emission from black holes.

It remains to include the one-loop contribution from the
reparametrization ghosts, dilaton and conformal mode.  If the number
of matter fields in the theory is large this contribution will be
insignificant compared to \sixth , which scales with $N$.  Therefore
our model, as it stands, at the very least provides a good
description of large $N$ black holes.  The most straightforward way
to include the ghosts, while preserving the symmetry \third , is to
shift the value of $\kappa$ in \sixth\ to ${N-24\over 12}$.  This has
the desirable feature that it turns the theory into a conformal field
theory with vanishing total central charge and thus makes it one-loop
finite [\jrat -\dealw].  However, the flux of Hawking radiation from
a black hole is proportional to $\kappa$ so this shift leads to the
undesirable result of unphysical modes contributing to the Hawking
radiation [\strom].

Strominger has suggested an alternate prescription for the one-loop
ghost contribution which is designed to decouple the ghosts from the
outgoing energy flux.  In our model his method boils down to keeping
$\kappa ={N\over 12}$ in \sixth\ and including the following term in
the action,
$$
S_{\rm ghost} = {26-2\over 12\pi} \int d^2x\>
\bigl[\partial _+(\rho -\phi)\partial _-(\rho-\phi)] \> .
\eqn\eighth
$$
Adding this term does not violate the symmetry \third .  In fact the
variation of $S_{\rm ghost}$ vanishes for all solutions of the
semi-classical equations of motion obtained from \sixth .  This is as
it should be since the role of the ghosts should not be to modify
equations of motion but rather to implement the gauge-fixing of the
path integral over off-shell geometries.  Unfortunately the theory is
no longer a conformal field theory with vanishing central charge if
this prescription is used but exact solutions can still be found.
These solutions exhibit reasonable physical behavior for all values
of $N$ and in particular the rate of Hawking evaporation is always
proportional to $N$.

In this paper we will not attempt to resolve all the issues involved
in the quantization of these models, but will focus on exact
semi-classical solutions and their physical properties.  We will
assume that $\kappa$ takes a positive value in the following
analysis.\foot{If we do not adopt Strominger's prescription but
rather perform the shift of $\kappa$ in \sixth , then $\kappa$ will
be negative for $N<24$.  In this case no singularity is encountered
in gravitational collapse, which might sound attractive, but as
pointed out above the system is unstable and emits negative energy
Hawking radiation.}  Our results therefore apply both to the $N>24$
conformal model \sixth\ (with $\kappa ={N-24\over 12}$) and to the
Strominger type theory with $\kappa ={N\over 12}$.

\chapter{Exact solutions}

To solve our model we follow Bilal and Callan [\bilcal] and de~Alwis
[\dealw] and perform a field redefinition to a Liouville theory.
Our new fields are defined as
$$\eqalign{
\Omega =& {\sqrt{\kappa}\over 2} \phi
+ {e^{-2\phi}\over \sqrt{\kappa}}\> , \cr
\chi =& \sqrt{\kappa} \rho - {\sqrt{\kappa}\over 2} \phi
+ {e^{-2\phi}\over \sqrt{\kappa}}\> . \cr}
\eqn\ninth
$$
The effective action \sixth\ takes the
form
$$
S = {1\over \pi} \int d^2x\>
\bigl[-\chip\chim + \omp\omm
+ \gl^2 e^{{2\over \sqrt{\kappa}}(\chi -\Omega)}
-{1\over 2}\sum_{i=1}^N \p_+ f_i \p_- f_i \bigr]\>
\eqn\tenth
$$
and the constraints become
$$
\kappa \, t_\pm = -\p_\pm\chi \p_\pm\chi
+\sqrt{\kappa}\, \p_\pm^2\chi
+\p_\pm\Omega \p_\pm\Omega
+ {1\over 2} \sum_{i=1}^N \p_\pm f_i \p_\pm f_i \> .
\eqn\eleventh
$$
Notice that $\Omega$ is bounded from below in \ninth\ and therefore
\tenth\ defines a rather unconventional quantum field theory.  In
this paper we only work with semi-classical equations and it should
be kept in mind that the full quantum theory may well describe very
different physics in regions of strong coupling.

The equations of motion derived from the Liouville action \tenth\ can
be solved exactly.  Let us first consider asymptotically flat static
geometries,
$$
\Omega =\chi =
- {\gl^2 x^+x^- \over \sqrt{\kappa}}
+P\sqrt{\kappa}\log{(-\gl^2 x^+x^-)}
+ {M\over \gl\sqrt{\kappa}} \> ,
\eqn\twelfth
$$
where $P$ and $M$ parametrize different solutions.  We are using
``Kruskal'' coordinates [\cghs,\rst] which is the coordinate system
where $\phi=\rho$.  Comparing with the definitions \ninth\
immediately reveals that the solution with $P=-{1\over 4}$ and $M=0$
is the familiar linear dilaton vacuum, $e^{-2\phi}=e^{-2\rho}=-\gl^2
x^+x^-$.

Adjusting the value of $P$ corresponds to having different energy
flux at spatial infinity in these solutions.  In particular,
solutions with $P=-{1\over 4}$ have vanishing asymptotic energy
density, whereas a geometry with $P=0$ has a smooth horizon (at
$x^+x^-=0$) and describes a quantum black hole in thermal
equilibrium with its environment [\bghs].  Solutions with $P\neq
-{1\over 4}$ have infinite ADM-mass because in all these solutions
there is non-vanishing energy density at infinity.  For the
$P=-{1\over 4}$ solutions the ADM-mass is $M$.  Static solutions with
positive ADM-mass are weakly coupled but singular at $x^+x^-=0$
[\bghs -\lslt].  Solutions with negative ADM-mass have a naked
singularity at a finite value of $-x^+x^-$.

The thermal equilibrium solutions (with $P=0$) have an interesting
property.  Two such static solutions with different values of the
parameter $M$ can be continuously matched across an infall line,
$x^+=x^+_0$, up to a uniform shift of $x^-$.  This corresponds to a
black hole absorbing an incoming shock wave with the shift in $M$
being equal to the energy carried by the wave.  The fact that the
solution remains static indicates that the Hawking temperature of
two-dimensional black holes remains independent of their mass, even
when our quantum corrections are added.  This can also be checked by
direct calculation using the geometry \twelfth .

Now let us consider a dynamical situation where an incoming shock
wave carries energy into the vacuum.  The corresponding solution is
constructed by patching together across an infall line the linear
dilaton solution and a time-dependent solution (with $P=-{1\over 4}$)
which describes the subsequent evolution of the black hole,
$$
\Omega =\chi = - {\gl^2 x^+x^- \over \sqrt{\kappa}}
-{\sqrt{\kappa}\over 4}\log{(-\gl^2 x^+x^-)}
- {m\over \gl\sqrt{\kappa}x^+_0}
(x^+{-}x^+_0)\theta (x^+{-}x^+_0) \> ,
\eqn\fourteenth
$$
The matching conditions at $x^+=x^+_0$ are provided by the $++$
constraints in \eleventh\ with ${1\over
2}\sum_{i=1}^N\p_+f_i\p_+f_i={m\over \gl x^+_0} \delta(x^+{-}x^+_0)$
where $m$ is the energy carried by the incoming shock wave.

The Liouville fields $\Omega$ and $\chi$ are non-singular in the
solution \fourteenth\ but in terms of the original variables, $\phi$
and $\rho$, a singularity forms on the infall line at $\phi =
\phi_{\rm cr}= {1\over 2} \log{\kappa \over 4}$.  This is easy to see
by computing, for example, the curvature scalar $R = 8 e^{-2\rho}
\rhopm$.  Using the transformations \ninth\ and the conservation
equation \fifth\ one finds
$$
\rhopm ={1\over \Omega '}\bigl[\chipm
-{\Omega ''\over \Omega '^2}\omp\omm \bigr] \> ,
\eqn\fifteenth
$$
The singularity forms where $\Omega '\equiv {d\Omega \over d\phi} =
{\sqrt{\kappa}\over 2} - {2\over \sqrt{\kappa}}e^{-2\phi}$ vanishes.
It lies
on a curve $(\bar x^+,\bar x^-)$ which is the constant $\phi$ contour
of \fourteenth\ at $\phi = \phi_{\rm cr}$ and is defined by the
following equation,
$$
1-\log{\kappa\over 4} =
-{4\gl^2\over \kappa}\bar x^+\bar x^-
-\log{(-\gl^2 \bar x^+\bar x^-)}
- {4m\over \gl\kappa x^+_0}
(\bar x^+{-}x^+_0)\,\theta (\bar x^+{-}x^+_0) \> .
\eqn\sixteenth
$$
This singularity occurs at the boundary of the range of $\Omega$
which is deep in the quantum mechanical strong coupling region.  It
may therefore well be absent in the full quantum theory.\foot{A shock
wave sent into a static geometry with positive ADM-mass (and
$P=-{1\over 4}$) will lead to topology change.  There is no
singularity on the infall line in this case but one will form at a
later time splitting the space into two disconnected regions.
Topology change of this sort will not occur if the initial state is
the linear dilaton vacuum.}

The singularity forms inside an apparent horizon, which is located
where $\phip=0$ [\rst].  The apparent horizon defines another curve
$(\hat x^+,\hat x^-)$ above the infall trajectory,
$$
\hat x^+ = -{\kappa \over 4\gl^2}\,
{1\over \hat x^- +{m\over \gl^3 x^+_0}}
\eqn\seventeenth
$$
When a massive black hole begins to evaporate the apparent horizon
recedes at a rate which agrees with calculations in the original
CGHS-model.   The agreement will hold as long as the remaining mass
is large compared to $N\gl$.  We were not able to follow the
evaporation to completion previously but this is straightforward now
that we have exact solutions.  In [\rst] we conjectured that the
singularity would always remain inside the apparent horizon and that
the geometry would have a global horizon separating the two.
However, the exact solution \fourteenth\ exhibits very different
behavior.  As suggested by Hawking [\haw] the singularity and the
apparent horizon collide in a finite proper time.  The intersection
point $(x^+,x^-)=(x^+_s,x^-_s)$ of the two curves \sixteenth\ and
\seventeenth\ is given by
$$\eqalign{
x^+_s=& {\kappa\gl x^+_0\over 4m}
(e^{4m\over \kappa\gl}-1) \> ,
\cr
x^-_s=& -{m\over \gl^3 x^+_0} \,
{1\over (1-e^{-{4m\over \kappa\gl}})} \> .
\cr}
\eqn\eighteenth
$$
The singularity goes from being space-like behind the apparent
horizon to being time-like, and therefore naked, after the two have
merged.  As a result the future evolution is not uniquely determined
unless boundary conditions are imposed at the naked singularity.

We emphasize once again that the naked singularity occurs deep in the
quantum mechanical region where the semi-classical theory is not
applicable.  The occurrence of such singularities means that very
quantum mechanical effects are in causal contact with outside
observers.  The precise nature of the phenomenon is beyond our
present knowledge.  It therefore seems appropriate to replace the
detailed dynamics of the naked singularity by phenomenological
boundary conditions.  In particular, we can expect radiation of
quantum mechanical origin out along the null line $x^-=x^-_s$.
Hawking has speculated that the emergence of the naked singularity
would precipitate a cataclysmic event, a thunderbolt, which
propagates outwards at the speed of light [\hawtalk].

\chapter{The final state}

A possible boundary condition which suggests itself is to
analytically continue the solution \fourteenth\ across the null line
$x^-=x^-_s$ from region~I into region~II in Figure~1.  This, however,
does not
lead to reasonable behavior in the asymptotic future.  To see this we
observe that all contours of constant $\phi =\phi_0 <\phi_{\rm cr}$
enter into region~II.  These contours are time-like everywhere
outside the apparent horizon and can be used to define fiducial
observers.  One finds that the scalar curvature tends to $-\infty$ as
$x^+\rightarrow\infty$ along every such a contour,
$$
R \sim -\kappa\gl^2\,
{e^{-4\phi_0}\over (e^{-2\phi_0}-e^{-2\phi_{\rm cr}})^3}
\log{\Bigl({x^+\over x^+_0}\Bigr)} \> .
\eqn\curvlim
$$
Thus all fiducial observers eventually find their way into a region
of diverging curvature, no matter how far away from the black hole
they set out.  Fortunately, this disastrous conclusion is by no means
inevitable.

A more reasonable possibility, suggested by Strominger [\andy], is
that the boundary conditions can be chosen in such a way that each
fiducial observer eventually tends to a region with vacuum behavior.
Miraculously, this type of boundary condition occurs naturally in the
exact solution \fourteenth .  Both $\phi$ and $\rho$ take vacuum
values on the null line $x^-=x^-_s$ dividing regions I and II.  This
means we can match the evaporating solution \fourteenth\ in region~I
onto a linear dilaton configuration in region~II, which is shifted
with respect to the original vacuum,
$$
e^{-2\phi}=e^{-2\rho}=-\gl^2 x^
+(x^- + {m\over \gl^3 x^+_0}) \> .
\eqn\nineteenth
$$
The fields are continuous at $x^-=x^-_s$ but their $x^-$-derivatives
are not.  Evaluating the $- -$ constraints at this null line one
finds a delta function contribution,
$$
{1\over 2}\sum_{i=1}^N\p_-f_i\p_-f_i
=-{\kappa\over 4}{(1-e^{-{4m\over \kappa\gl}})
\over (x^- + {m\over \gl^3 x^+_0})}
\, \delta(x^--x^-_s) \> .
\eqn\twentieth
$$
This means that a matter shock wave carries a small amount
$-{\kappa\gl\over 4}(1-e^{-{4m\over \kappa\gl}})$ of negative energy
out along the line to null-infinity.  This result can be checked
independently as follows.  The energy carried by a black hole across
null lines of constant $x^-$ is given by
$$
m(x^-) =
\lim_{x^+\rightarrow\infty} {1\over 4\gl}
\,e^{-2\phi}\, R \>.
\eqn\twentyfirst
$$
This gives the correct ADM-mass for the $P=-{1\over 4}$ static
solutions and provides a definition of the remaining mass of an
evaporating black hole [\lslt].\foot{The definition in [\lslt]
included a factor of $(1-{N\over12}e^{2\phi})^{3\over 2}$ which goes
to $1$ as $x^+\rightarrow\infty$.}  Using the solution \fourteenth\
to evaluate the mass at $x^-=x^-_s$ gives precisely the same small
negative energy as found above.  When the negative energy shock wave
reaches null infinity it brings the energy to zero, the vacuum value,
in region~II.

The fact that negative energy is carried out from the naked
singularity looks strange but is not very serious.  First of all,
energy density is not positive definite in quantum theory and global
energy positivity is not violated.  Second, the amount of negative
energy is limited by $-{\kappa\gl\over 4}$ for all values of the
original black hole mass $m$ (and vanishes in the $m\rightarrow 0$
limit). The limiting value is the analog of the Planck scale in this
theory so this negative energy may simply be an artifact of our
semi-classical approximation.  At any rate the outgoing shock wave
does not represent a violent event on an astronomical scale so we
will refer to it as a thunderpop rather than a thunderbolt.

We find it rather compelling that it is possible to match the
evaporating solution onto the vacuum, with only a Planck mass worth
of adjustment needed to the energy.  The physical picture this
presents is that the black hole evaporates completely, leaving no
remnant behind.  The region $\phi > \phi_{\rm cr}$ is completely
unphysical in this case [\rst,\lslt,\bghs].  The vacuum should be
taken to be the linear dilaton solution for $\phi < \phi_{\rm cr}$
and some boundary conditions supplied at the critical line where it
is time-like.  Perhaps the appropriate framework for this system is
to couple two-dimensional gravity to non-trivial boundary degrees of
freedom, as considered in [\cgclt] in the context of open string
theory.

\chapter{General distributions of incoming matter}

Finally we want to consider arbitrary distributions of incoming
matter.  We will need expressions for the incoming flux of energy in
terms of Kruskal coordinates.  Let $\gl\sigma^\pm=\pm \log{(\pm
\gl x^\pm)}$.  In this coordinate system the metric asymptotically
approaches the Minkowski metric.  The energy is by definition
conjugate to ${1\over 2}(\sigma^++\sigma^-)$.  Transforming to
Kruskal coordinates is straightforward and the total energy of a
distribution of incoming matter is given in terms of the Kruskal
energy momentum tensor by
$$
M = \gl \int_0^\infty dx^+_0 \, x^+_0 \, T_{++}(x^+_0) \> .
\eqn\twentysecond
$$
Another quantity of interest is the total incoming Kruskal momentum
conjugate to $x^+_0$.  This is given by
$$
P_+ = \int_0^\infty dx^+_0 \, T_{++}(x^+_0) \> .
\eqn\twentythird
$$
We will also define $x^+$-dependent truncated versions of
\twentysecond\ and \twentythird\ as follows,
$$\eqalign{
M(x^+) =& \gl \int_0^{x^+} dx^+_0 \,
x^+_0 \, T_{++}(x^+_0) \> ,
\cr
P_+(x^+) =& \int_0^{x^+} dx^+_0 \, T_{++}(x^+_0) \> .
\cr}
\eqn\twentyfourth
$$
The exact solution which generalizes \fourteenth\ is
$$
\Omega =\chi =
- {\gl^2\over \sqrt{\kappa}}\,
x^+\bigl(x^-+{P_+(x^+)\over \gl^2}\bigr)
+ {M(x^+)\over \sqrt{\kappa}\gl}
-{\sqrt{\kappa}\over 4}\log{(-\gl^2 x^+x^-)} \> ,
\eqn\twentyfifth
$$
This solution will typically have naked singularities.  In this case
\twentyfifth\ is only applicable in those regions which are not in
the causal future of such singularities and we have to supply
additional boundary conditions.  We do not know what is the
``correct'' set of conditions to impose at a naked singularity but
for the purpose of illustration we will adopt a particularly simple
boundary condition, demanding that matter energy, carried by the
$f$-fields, is totally reflected from the $\phi =\phi_{\rm cr}$ line
where it is time-like,
$$
f_i = 0 \big\vert _{\phi =\phi_{\rm cr}} \> .
\eqn\twentysixth
$$
It is not meaningful to apply boundary conditions where the
singularity is space-like.

We consider first a simple example in which the incoming energy flux
is turned on at some finite time and remains steady at smaller rate
than the Hawking flux for a two-dimensional black hole.  In this case
the incoming energy momentum in Kruskal coordinates is
$$
T_{++}(x^+) = {\epsilon\over \gl {x^+}^2} \,
\theta(x^+{-}x^+_0) \> ,
\eqn\twentyseventh
$$
where $\epsilon < {\kappa\gl\over 4}$ is the constant energy flux and
$x^+=x^+_0$ defines the leading edge of the incoming energy.  The
solution \twentyfifth\ reduces to
$$
\Omega =\chi =
- {\gl^2\over \sqrt{\kappa}}\,
x^+(x^-+{\epsilon\over \gl^3x^+_0})
-{\sqrt{\kappa}\over 4}\log{(-\gl^2 x^+x^-)} \>
+ {\epsilon\over \sqrt{\kappa}\gl}\,
(1+\log{x^+\over x^+_0}) \> .
\eqn\twentyeighth
$$
The curve $\phi =\phi_{\rm cr}$ is time-like for $\epsilon <
{\kappa\gl\over 4}$, as shown in Figure~2.  Region~i, where $x^- <
-{\kappa\over 4\gl^2x^+_0}$, is not in causal contact with any
singularity.  On the other hand region~ii, where $x^- > -{\kappa\over
4\gl^2x^+_0}$, can receive signals from the singularity and therefore
the solution \twentyeighth\ is not correct in this region.

The simple reflecting boundary conditions \twentysixth\ imply the
following relation between the incoming and outgoing values of the
matter energy momentum,
$$
T_{--}^f(x^-)=
T_{++}^f(\bar x^+)\,
\bigl({\partial \bar x^+\over \partial x^-}\bigr)^2 \> ,
\eqn\twentyninth
$$
where $x^+=\bar x^+(x^-)$ defines the boundary and
$T_{\pm \pm}^f={1\over 2}\sum_{i=1}^N \p_\pm f_i \p_\pm f_i$.  These
boundary conditions along with the equations of motion are sufficient
to completely determine the fields as well as the boundary curve
itself in region~ii in Figure~2.  In our simple example of a small
uniform incoming flux the solution in region~ii turns out to be a
static configuration \twelfth\ with $x^-$ shifted by ${\epsilon\over
\gl^3x^+_0}$ and parameters
$$\eqalign{
P=&-{1\over 4}+{\epsilon\over \kappa\gl} \> ,
\cr
M=&\epsilon \, (1{-}\log{\kappa\over 4})
+({\kappa\gl\over 4}{-}\epsilon)
\log{(1{-}{4\epsilon\over \kappa\gl})} \> .
\cr}
\eqn\moments
$$
The solutions on either side of the line $x^-=-{\kappa\over
4\gl^2x^+_0}$ match smoothly across it.  In particular, there is no
outgoing shock wave propagating along this null line, as is easily
seen be evaluating the $--$ constraints there.

For a more general incoming flux distribution, which is smaller than
${\kappa\gl\over 4}$ at any given time, the solution in region~II is
more complicated but it can be constructed in terms of the incoming
flux distribution given the boundary conditions \twentysixth .  In
this case all information is reflected off the boundary in
the semi-classical approximation.  This can be interpreted as saying
that information loss does not occur in low energy (sub-planckian)
physics, at least at the semi-classical level.\foot{Note that a
classical shock wave cannot provide a good description of an incoming
low-energy state because of the uncertainty
principle, which ensures that a given
energy $\delta$ cannot be localized within a distance less than
${1\over \delta}$.}  This does not preclude the possibility of
information loss of a more quantum mechanical nature, {\it e.g.}
tunneling, in low energy processes.

The situation is very different when one considers an energy flux
$\epsilon > {\kappa\gl\over 4}$ for which the singularity goes
space-like and an actual black hole is formed.  This case is
qualitatively similar to the incoming shock wave considered in
section~2.  An apparent horizon forms.  If the flux is turned off at
some point the apparent horizon will recede and eventually collide
with the singularity, sending off a thunderpop.  The final state will
be the linear dilaton vacuum as before.  The boundary condition
we applied in section~2 is in fact a special case of the
reflecting boundary conditions we employed in this section.

Let us assume that the incoming flux is large enough from the start
for an apparent horizon to form and that it is maintained for a
finite
length of time.  In this case the solution \twentyfifth\ is valid
until the naked singularity emerges.  In particular, the bulk of the
outgoing Hawking radiation is found in the region to the right of an
outgoing null line analogous to $x^-=x^-_s$ in Figure~1 and therefore
described by \twentyfifth .  It is striking that the final
($x^+\rightarrow\infty$) behavior of the fields in that region
depends on only two moments, $M$ and $P_+$, of the incoming $T_{++}$
and not on the detailed history of the initial state.  Evidently most
of the information contained in the initial state is lost in this
one-loop semi-classical approximation.  It should however be noted
that the present situation is somewhat better than that without any
back-reaction because the classical no-hair theorem implies that the
final
state can only depend on one of these moments, {\it i.e.} the total
mass $M$.  It is tempting to conjecture that a more systematic
quantum treatment of the problem (for example including higher
gravitational loops) will introduce a dependence on higher moments,
$$
P^n_+ = \int_0^\infty dx^+_0 \, (x^+_0)^{-n+1}
\, T_{++}(x^+_0)  \> .
\eqn\thirtieth
$$
A simple electrodynamic system where unitarity is restored by quantum
corrections was considered in [\pst].

\noindent
{\undertext{Acknowledgements:}}  It is a pleasure to thank
A.~Strominger for many insightful suggestions and for comments on an
early manuscript of this paper.  We would also like to thank
A.~Bilal, C.~Callan and A.~Tseytlin for useful discussions.

\FIG\figone{Black hole formed by an incoming shock wave at
$x^+=x^+_0$.  A space-like singularity forms inside an apparent
horizon, which recedes until it collides with the singularity at
$x_s$.  The solution in region~II is determined by boundary
conditions at the naked singularity.}
\FIG\figtwo{A small steady incoming energy flux at $x^+=x^+_0$ leads
to a time-like singularity.  Region~i is not in causal contact with
the singularity but the solution in region~ii depends on the boundary
conditions imposed at $\phi=\phi_{\rm cr}$.}

\refout
\figout
\end